# Design of microring resonators integrated with 2D graphene oxide films for four-wave mixing

Yuning Zhang, Jiayang Wu, *Member, IEEE*, Yang Qu, Linnan Jia, Baohua Jia, *Fellow, OSA*
and David J. Moss, *Fellow, IEEE, Fellow, OSA*

*Abstract*—We theoretically investigate and optimize the performance of four-wave mixing (FWM) in microring resonators (MRRs) integrated with two-dimensional (2D) layered graphene oxide (GO) films. Owing to the interaction between the MRRs and the highly nonlinear GO films as well as to the resonant enhancement effect, the FWM efficiency in GO-coated MRRs can be significantly improved. Based on previous experiments, we perform detailed analysis for the influence of the GO film parameters and MRR coupling strength on the FWM conversion efficiency (CE) of the hybrid MRRs. By optimizing the device parameters to balance the trade-off between the Kerr nonlinearity and loss, we achieve a high CE enhancement of ~18.6 dB relative to the uncoated MRR, which is ~8.3 dB higher than previous experimental results. The influence of photo-thermal changes in the GO films as well as variations in the MRR parameters such as the ring radius and waveguide dispersion on the FWM performance is also discussed. These results highlight the significantly improved FWM performance that can be achieved in MRRs incorporating GO films and provide a guide for optimizing their FWM performance.

## I. INTRODUCTION

Graphene oxide (GO) has become a rising star in the family of two-dimensional (2D) materials owing to its potential for mass production as well as the flexibility in tuning its material properties [1-4]. Recently, the excellent nonlinear optical properties of GO have attracted significant interest [5-9]. It has been reported that GO has an ultrahigh Kerr nonlinearity ($n_2$) that is about 4 orders of magnitude higher than nonlinear bulk materials such as silicon and chalcogenide glasses [5, 6, 10]. In addition, GO has a large optical bandgap (typically > 2 eV [1, 11]), which yields a material absorption that is over 2 orders of magnitude lower than graphene as well as negligible two-photon absorption (TPA) in the telecom band [12, 13]. Another important advantage of GO is that it can be mass produced from natural graphite [3]. This, together with facile solution-based fabrication processes [14], is attractive for large-scale manufacturing of integrated devices that incorporate GO films [2, 15, 16].

Based on these advantages, many high performance nonlinear photonic devices that incorporate GO films [13, 17-21] have been demonstrated – especially those based on complementary metal-oxide-semiconductor (CMOS) compatible integrated platforms [13, 17-19]. Enhanced four-wave mixing (FWM) in GO-coated doped silica and silicon nitride (SiN) waveguides has been reported [13, 17], where conversion efficiency (CE) enhancements of up to 6.9 dB and 9.1 dB relative to the uncoated waveguides were achieved. Significant spectral broadening of optical pulses in GO-coated silicon waveguides induced by self-phase modulation (SPM) has also been observed [19], achieving a high spectral broadening factor of 4.34 for a device with a patterned film including 10 layers of GO. A significant enhancement in the nonlinear figure of merit (FOM) for silicon nanowires by a factor of 20 was also achieved, resulting in a FOM > 5.

In our previous work [18], we experimentally demonstrated enhanced FWM in CMOS compatible doped silica microring resonators (MRRs) integrated with 2D layered GO films. Due to the resonant enhancement effect [22, 23], an increase of up to ~10.3 dB in the FWM CE was achieved. In this paper, we fully investigate and optimize the FWM performance of GO-coated MRRs based on previous experimental measurements of the GO film properties such as loss and Kerr nonlinearity, which are distinct from bulk materials and show a strong dependence on the film thickness and optical power. We perform a detailed analysis of the influence of the GO film parameters and MRR coupling strength on the FWM CE of the hybrid MRRs. By properly balancing the trade-off between the Kerr nonlinearity and loss, a high CE enhancement of ~18.6 dB relative to the uncoated MRR is achieved, which is ~8.3 dB higher than what has been achieved experimentally. We also discuss the influence of photo-thermal changes in the GO films as well as the variation of other MRR parameters such as ring radius and waveguide dispersion on the FWM performance. These results highlight the significant potential to improve on previous experimental results [18] and provide a guide for optimizing FWM performance of MRRs integrated with GO films.

This work was supported by the Australian Research Council Discovery Projects Programs (No. DP150102972 and DP190103186), the Swinburne ECR-SUPRA program, the Industrial Transformation Training Centers scheme (Grant No. IC180100005), and the Beijing Natural Science Foundation (No. Z180007). *(Corresponding author: Jiayang Wu, Baohua Jia, and David J. Moss)*

Y. N. Zhang, J. Y. Wu, Y. Qu, L. N. Jia, and D. J. Moss are with the Optical Sciences Center, Swinburne University of Technology, Hawthorn, VIC 3122, Australia. (e-mail: yuningzhang@swin.edu.au, jiayangwu@swin.edu.au, yqu@swin.edu.au, ljia@swin.edu.au, dmoss@swin.edu.au).

B. H. Jia is with Center for Translational Atomaterials, Swinburne University of Technology, Hawthorn, VIC 3122, Australia. (e-mail: bjia@swin.edu.au)

## II. DEVICE STRUCTURE

Fig. 1(a) shows a schematic of an integrated MRR made from doped silica, with 1 layer of patterned GO film being coated on the planarized waveguide top surface. Inset shows a schematic illustration for the atomic structure of GO, including different oxygen-containing functional groups (OFGs) such as hydroxyl, epoxide, and carboxylic decorated on a graphene-like carbon lattice. In contrast to graphene, which has a metallic behavior with a zero bandgap [24], pristine GO is a dielectric material with a bandgap > 2 eV [1, 12]. This is larger than both the single photon (~0.8 eV) and two-photon (~1.6 eV) energies around 1550 nm, resulting in negligible linear light absorption or TPA in the telecom band. We consider MRRs that are fabricated on a high index doped silica glass (Hydex) platform [25] via CMOS compatible processes. The bending loss of the doped silica MRRs studied here, with a radius of 592 μm and a core-cladding index contrast of 17%, is negligible. According to Ref. [22] and references therein, this index contrast is sufficient to support MRRs down to a minimum radius of ~20 μm. More details about the Hydex device fabrication can be found in Refs. [22, 26, 27]. As compared with GO-coated waveguides, GO-coated MRRs can dramatically improve the FWM efficiency by virtue of the resonant enhancement of the optical intensity within the resonant cavities [22, 23], thus significantly reducing the required power. The upper cladding of the doped silica MRR is removed by chemical mechanical polishing (CMP) to obtain a planarized waveguide top surface for GO film coating. The GO film coating can be achieved using a solution-based method that yields layer-by-layer film deposition and precise control of the film thickness with an ultrahigh resolution of ~2 nm [12, 28]. Unlike graphene or other 2D materials that are typically prepared via non-solution-based deposition followed by cumbersome layer transfer processes [29-32], our coating method enables large-area, transfer-free, and high-quality GO film coating with high fabrication stability, mass producibility, and excellent film attachment onto integrated waveguides [2, 19]. Patterning of the films can be achieved via standard lithography and lift-off processes [18, 28]. This, together with the layer-by-layer deposition of GO films, forms the basis for the optimization of the FWM performance of the hybrid MRRs with different GO film thicknesses and pattern lengths.

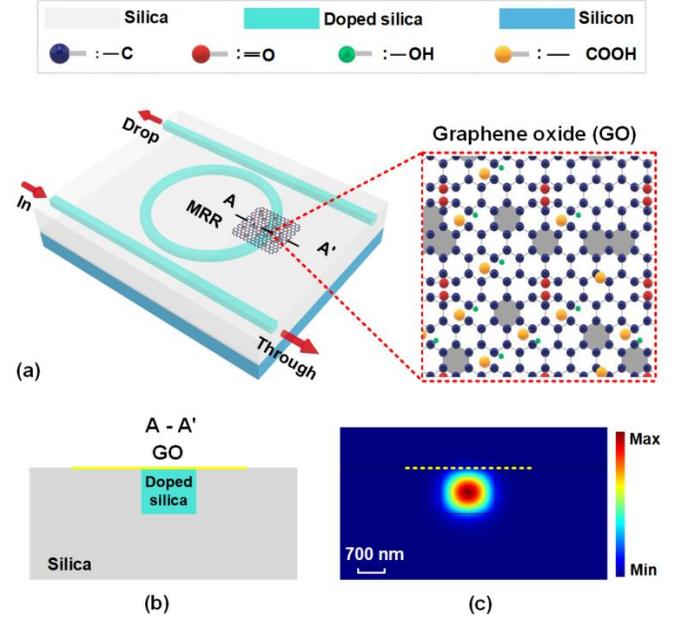

Fig. 1. (a) Schematic illustration of an integrated doped silica MRR coated with 1 layer of patterned GO film. Inset shows a schematic of atomic structure of GO. (b) Schematic illustration of the cross section of the hybrid MRR in (a). (c) TE mode profile corresponding to (b).

TABLE I
PARAMETERS OF DOPED SILICA MRR, GO FILM, AND CW LASER

| | | Refractive index [a] | Extinction coefficient | Kerr coefficient (m$^2$/W) |
|---|---|---|---|---|
| Doped Silica MRR | Material parameters | $n_{hydex}$ : 1.7 [25] | $k_{hydex}$ : 0 [18] | $n_{2\text{-}hydex}$ : 1.3 × 10$^{-19}$ [18] |
| | | Transmission / coupling coefficients | Radius [c] | Propagation loss (dB/cm) |
| | Physical parameters | $t, \kappa$ [b] | $R$ | 0.25 [18] |
| GO film | Material parameters | Refractive index | Extinction coefficient [d] | Kerr coefficient (m$^2$/W) |
| | | $n_{GO}$ : 1.97 [13] | $k_{GO}(N)$ : 0.0074 – 0.0189 [18] | $n_{2\text{-}GO}(N)$ : 1.7 × 10$^{-14}$ – 2.7 × 10$^{-14}$ [18] |
| | Physical parameters | Thickness for 1 layer [e] | GO layer number | Coating length |
| | | $d$ : 2 nm [18] | $N$ | $L_c$ |
| CW laser | Physical parameters | CW power for loss measurement | Pump power for FWM | Signal power for FWM |
| | | $P_{CW}$ | $P_p$ | $P_s$ |

[a] Here we show the refractive indices at 1550 nm, the same applies for other material parameters in this Table.
[b] $t^2 + \kappa^2 = 1$ for lossless coupling is assumed for the directional couplers.
[c] The circumference of the MRR is $L = 2\pi R$.
[d] Here we show the extinction coefficient and Kerr coefficient at $P_{CW}$ = 25 dBm for $N$ = 1 – 50 based on the measured results in Ref. [18].
[e] Following our previous experimental measurements [18], the GO film thickness is assumed to be proportional to $N$, with a thickness of 2 nm per layer.

Fig. 1(b) shows a schematic of the waveguide cross section of the hybrid MRR in Fig. 1(a). The corresponding transverse electric (TE) mode profile is shown in Fig. 1(c). We chose the TE polarization in our following analysis because it supports an in-plane interaction between the film and the evanescent field leaking from the MRR, which is much stronger than the out-of-plane interaction due to the significant optical anisotropy of 2D films [28, 31, 32]. Table I summarizes the parameters of the doped silica MRR, the GO film, and the continuous-wave (CW) laser used in our following analysis, with the former two being further classified into material and physical parameters. Four-port MRRs with two identical directional couplers are used in our following analysis, which is consistent with that used in Ref. [18]. The GO-coated MRRs are designed based on, but not limited to, the Hydex platform. We have reported a significant enhancement of the nonlinear optical performance in both GO-coated SiN and silicon waveguides [17, 19]. The investigation of the nonlinear optical performance of GO-coated SiN and silicon MRRs will be the subject of future work.

In the following sections, we first investigate the power-dependent Q factors, propagation loss, and nonlinear parameters of the hybrid MRRs induced by photo-thermal changes in the GO films. Next, by properly balancing the trade-off between loss and the Kerr nonlinearity, we optimize the FWM CE in the hybrid MRRs by regulating the GO film parameters ($N$, $L_c$) and the MRR coupling strength ($t$). Finally, we discuss the influence of photo-thermal changes in the GO films as well as the effect of varying other MRR parameters such as the ring radius and waveguide dispersion on the FWM performance of the hybrid MRRs.

## III. POWER-DEPENDENT PROPAGATION LOSS AND NONLINEAR PARAMETERS

As reported in previous work [18], the linear loss ($k$) and Kerr nonlinearity ($n_2$) of GO films coated on integrated waveguides change with input CW power, particularly at high powers. This can be attributed to photo-thermal changes in the GO films, which is a combined result of power sensitive photo-thermal reduction as well as self-heating and thermal dissipation in the multilayered film structure [17, 18, 33]. Such changes are not permanent and can revert back after the power is turned off. In addition, these changes have a slow time response on the order of millisecond, which is different to FWM and TPA that have ultrafast response times on the order of femtoseconds [17]. Photo-thermal changes in the GO films lead to power-dependent propagation loss and nonlinear parameters for GO-coated waveguides, and this is further amplified in GO-coated MRRs due to resonant enhancement. In this section, we investigate the power-dependent Q factors, propagation loss, and nonlinear parameters of the hybrid MRRs induced by the photo-thermal changes in GO films.

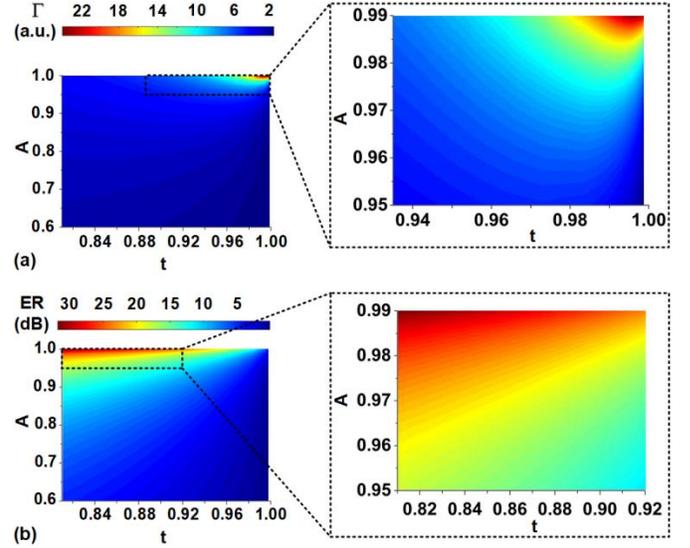

Fig. 2. (a) $\Gamma$ versus $t$ and $A$. (b) ER versus $t$ and $A$. The insets in (a) and (b) show the corresponding zoom-in views.

We first calculate the resonant build-up factor ($\Gamma$) of a MRR as a function of its coupling strength ($t$) and round-trip field transmission factor ($A$). The $\Gamma$ reflects the relationship between the input CW power ($P_{CW}$) and the intracavity power ($P_{intra}$) in a MRR, which will be used for calculating $P_{intra}$ directly related to the propagation loss and nonlinear linear parameters of the hybrid MRRs in our following analysis. Fig. 2(a) shows $\Gamma$ versus $t$ and $A$. The $\Gamma$ was calculated at resonant wavelengths based on [34, 35]:

$$\Gamma = \frac{P_{intra}}{P_{CW}} = (1-t^2)t^2 A^2 / (1-2t^2 A + t^4 A^2) \quad (1)$$

In Eq. (1), $A$ can be further expressed as:

$$A = \exp\left(-\tfrac{1}{2}\alpha_u L_u\right)\exp\left(-\tfrac{1}{2}\alpha_c L_c\right) \quad (2)$$

where $\alpha_{c,u}$ and $L_{c,u}$ are the loss factors and lengths of the GO coated and uncoated waveguide segments, respectively. Since $\alpha_c$ (e.g., 1.27 dB/cm for $N$ = 1 and 29.43 dB/cm for $N$ = 10) is much higher than $\alpha_u$ (i.e., 0.25 dB/cm), $A$ in Eq. (2) is mainly determined by $L_c$. In Fig. 2(a), the maximum $A$ is 0.989, which corresponds to the uncoated MRR. Unless otherwise specified, the MRR radius used is 592 μm – the same as in Ref. [18]. We choose such a MRR because it has a long circumference, thus providing a large range to adjust the GO coating length for optimizing the FWM performance. The maximum $\Gamma$ is achieved at $t$ = 0.994 and $A$ = 0.989, which is determined by the balance between $t$ and $A$, as reflected by Eq. (1). Fig. 2(b) shows the MRR's extinction ratio (ER) versus $t$ and $A$. The ER is defined as the resonance notch depth at the through port. The ER increases with $A$ but decreases with $t$, mainly due to the change in the difference between intracavity loss and external coupling loss of a four-port MRR with two identical directional couplers [36, 37].

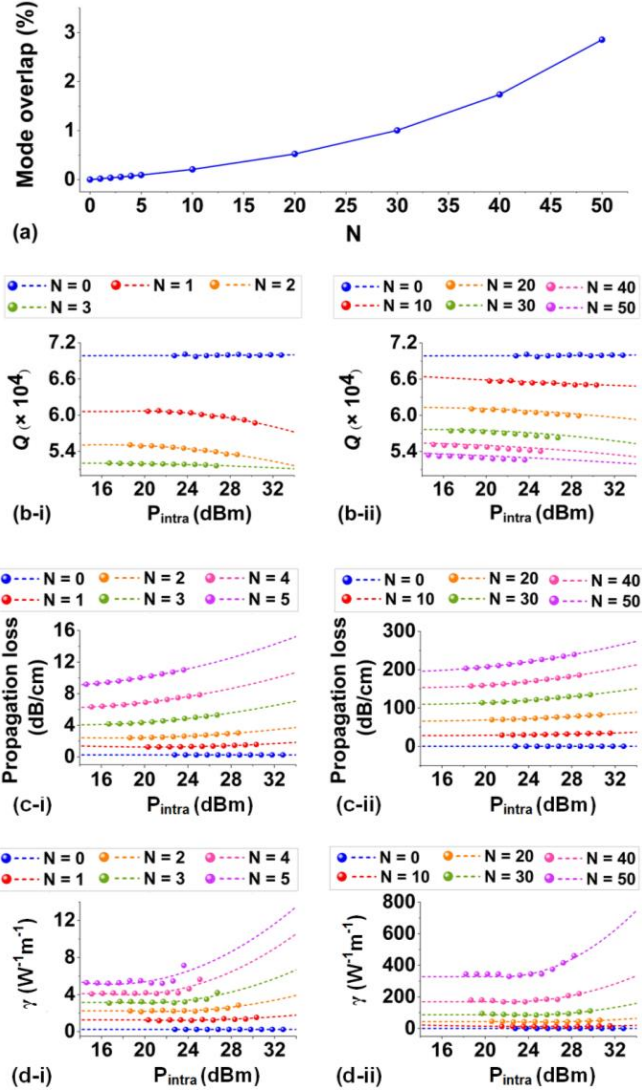

Fig. 3. (a) GO mode overlap versus GO layer number $N$. (b)−(d) Fit Q factors, propagation loss and nonlinear parameters $\gamma$ of GO-coated MRRs versus intracavity power $P_{intra}$ based on previous measured results for hybrid MRRs with (i) 1 − 5 layers of uniformly coated and (ii) 10 − 50 layers of patterned GO, respectively. The result for the uncoated MRR ($N$ = 0) is also shown for comparison.

Fig. 3(a) shows GO mode overlap in the GO-coated doped silica waveguides versus layer number $N$. Most light is confined within the waveguide core and only a small portion (< 3% for $N$ = 50) overlaps with the GO films, mainly resulting from the large difference in their cross-sectional areas. In previous work [18], we measured the Q factors, propagation loss, and nonlinear parameters versus input CW power ($P_{CW}$) for hybrid MRRs with 1 − 5 layers of uniformly coated and 10 − 50 layers of patterned (50-μm-long) GO, respectively. The coupling strength of the uncoated MRR was 0.912. In Figs. 3(b) − (d), we fit the measured power-dependent Q factors, propagation loss, and nonlinear parameters as functions of the intracavity power $P_{intra}$, which will be used for calculating FWM CE in next section. Note that the propagation loss and nonlinear parameters are the average values for the GO coated segments (considering the power-dependent photo-thermal changes). The input CW power $P_{CW}$ in Ref. [18] (from 15 dBm to 25 dBm) is converted to corresponding intracavity power $P_{intra}$ based on the calculated $\Gamma$ in Fig. 2(a). In Figs. 3(b) and (c), the Q factor decreases with GO layer number $N$ while the propagation loss shows an opposite trend. This is mainly due to an increase in GO mode overlap for the hybrid MRRs with thicker GO films. A small contribution is from an increase in the material absorption arising from inhomogeneous defects and imperfect contact between the multiple GO layers [13, 28]. As $P_{intra}$ increases, the hybrid MRRs show decreased Q factors and increased propagation loss, in contrast to the uncoated MRR that manifests a constant Q factor and propagation loss. This further confirms the power sensitive photo-thermal changes in GO films. Following the same trends with the propagation loss, the nonlinear parameter $\gamma$ in Fig. 3(d) increases with both $N$ and $P_{intra}$. This reflects the trade-off between the Kerr nonlinearity and linear loss, which is critical for optimizing the FWM performance. Note that in our calculation we neglect the influence of power-dependent loss on the round-trip field transmission factor $A$, since accounting for it would only lead to a maximum difference in $\Gamma < 0.7\%$.

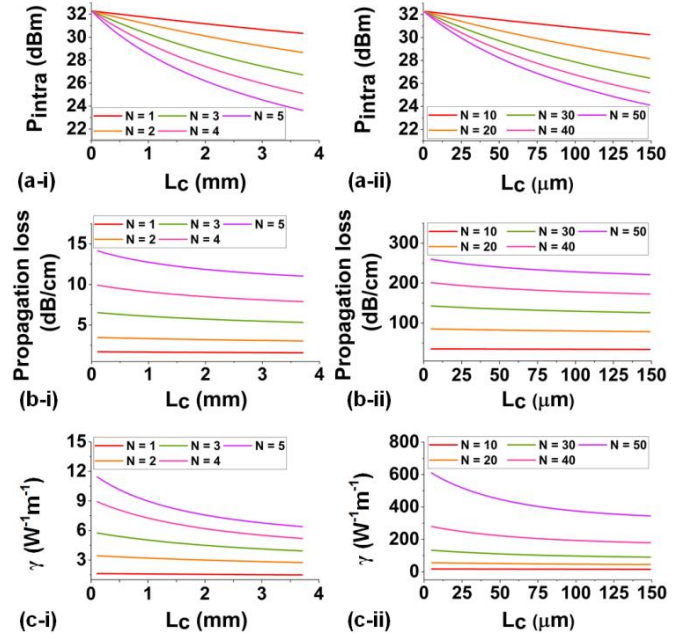

Fig. 4. (a) Intracavity power $P_{intra}$, (b) propagation loss, and (c) nonlinear parameters $\gamma$ versus coating length $L_c$ for the hybrid MRRs with films including (i) $N$ = 1 – 5 and (ii) $N$ = 10 – 50 GO layers, respectively. In (a) – (c), $t$ = 0.912, $R$ = 592 μm, and $P_{CW}$ = 25 dBm.

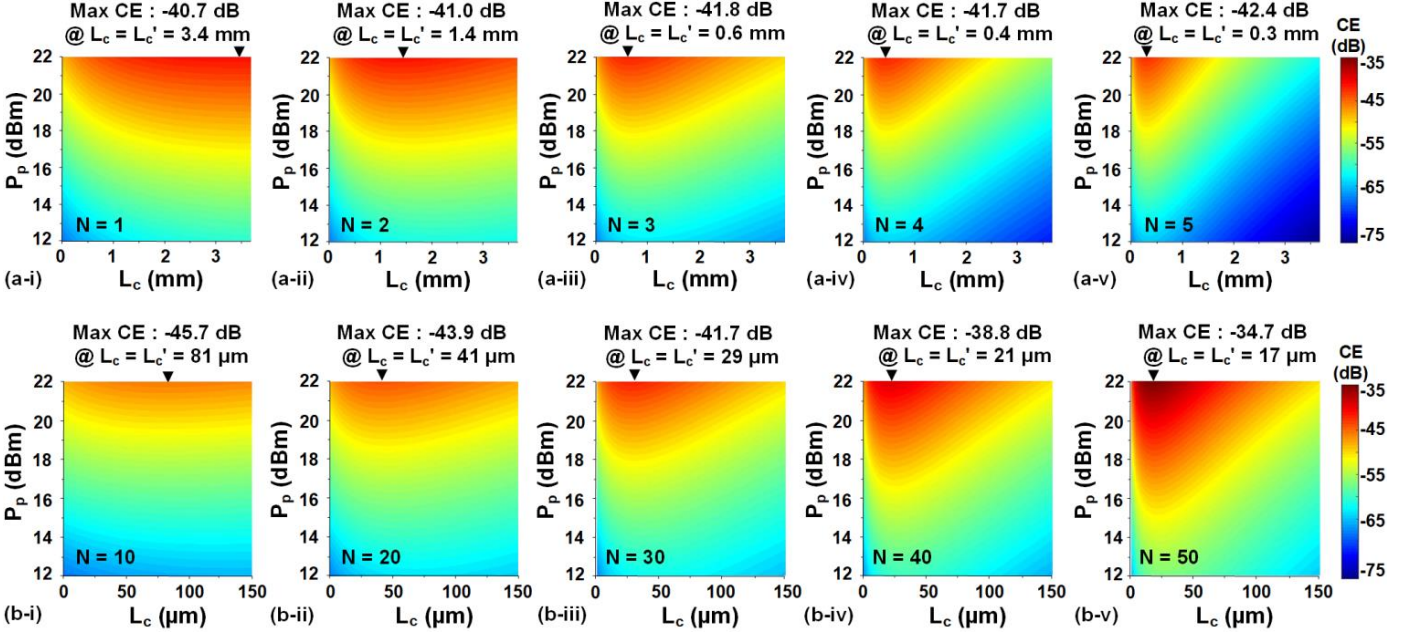

Fig. 5. FWM CE of GO-coated MRRs versus $L_c$ and $P_p$ when (a) $N$ = 1 − 5 and (b) $N$ = 10 − 50. In (a) and (b), $t$ = 0.912, $R$ = 592 μm. The corresponding result for the uncoated MRR (when $L_c$ = 0 mm) is also shown.

TABLE II
PROCESS FLOW TO CALCULATE FWM CE IN GO-COATED MRRS

| Pre-step [a] | Aim | Method & Theory | Used parameters |
|---|---|---|---|
| 0 | Fit $\gamma(P_{intra})$, $\alpha_c(P_{intra})$ based on the experimental results in Ref. [18] | Matlab | Measured $\gamma$ and $\alpha_c$ in Ref. [18] |
| Step | Aim | Method & Theory | Used parameters |
| 1 | Calculate round-trip transmission $A(N, L_c, \alpha_c)$ | Eq. (2) | $N, L_c, \alpha_c$ |
| 2 | Calculate build-up factor $\Gamma(A, t)$ | Eq. (1) | $t$ and results in Step 1 |
| 3 | Calculate intracavity power $P_{intra}(\Gamma, P_p, P_s)$ | Eq. (1) | $P_p, P_s$, and results in Step 2 |
| 4 | Calculate $\gamma(P_{intra})$ and $\alpha_c(P_{intra})$ | Matlab | Results in Pre-Step and Step 3 |
| 5 | Calculate CE of GO-coated MRRs [b] | Matlab Eqs. (3) – (5) | $t, \kappa$, and results in Step 1, 4 |

[a] The pre-step is done in Fig. 3 of Section III.
[b] To optimize FWM CE, Step 1 – 5 were repeated to calculate the CEs for the hybrid MRRs with various $N$, $L_c$, and $t$.

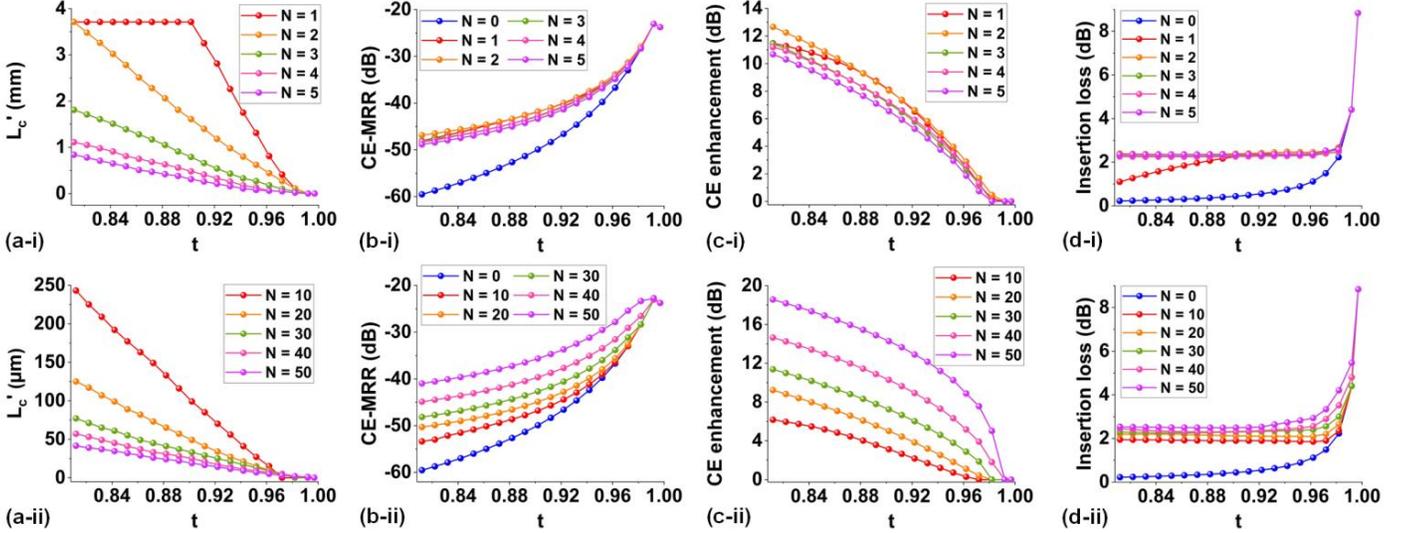

Fig. 6. (a) Optimized coating length $L_c'$ versus $t$. (b) CE of hybrid MRRs with the optimized coating length $L_c'$ in (a). (c) CE enhancement of the hybrid MRRs extracted from (b). (d) Insertion loss of the hybrid MRRs with the optimized coating lengths $L_c'$ in (a). In (a) – (d), $R = 592$ μm, $P_p = P_s = 22$ dBm, (i) and (ii) show the results for $N = 1 - 5$ and $N = 10 - 50$, respectively. In (b) and (d), the results for the uncoated MRRs ($N = 0$) are also shown for comparison.

For a fixed input power $P_{CW}$, varying the GO film parameters such as layer number $N$ and coating length $L_c$ changes the intracavity loss and hence intracavity power $P_{intra}$. Therefore, the power dependent propagation loss and nonlinear parameters of the hybrid MRRs are also affected by $N$ and $L_c$. Fig. 4(a) shows $P_{intra}$ versus $L_c$ for the hybrid MRRs with films including (i) $1 - 5$ and (ii) $10 - 50$ layers of GO. The other parameters are kept constant: $t = 0.912$ and $P_{CW} = 25$ dBm – taken from our previous experiments [18]. To clearly show the difference, we choose different ranges for $L_c$ in Figs. 4(a-i) and (a-ii) – with a smaller range for thicker films ($N \geq 10$). As can be seen, $P_{intra}$ decreases with $L_c$ and $N$, resulting from an increased intracavity loss in the hybrid MRRs. Figs. 4(b) and (c) show the corresponding propagation loss and nonlinear parameters $\gamma$ versus $L_c$, respectively. Both the propagation loss and nonlinear parameters $\gamma$ decrease with $L_c$, showing a trend similar to that of $P_{intra}$ in Fig. 4(a) and reflecting that the power dependent propagation loss and nonlinear parameters of the hybrid MRRs is strongly dependent on $P_{intra}$.

TABLE III
COMPARISON OF HYBRID MRRS WITH OPTMIZED GO COATING LENGTHS AND THOSE IN PREVIOUS EXPERIMENT

| $N$ | Experimental results [a] | | | | Optimized $L_c'$ for fixed $t = 0.912$ | | | |
|---|---|---|---|---|---|---|---|---|
| | $L_c$ (mm) | $t$ | Max CE (dB) | Max CE enhancement (dB) | $L_c'$ (mm) | $t$ | Max CE (dB) | Max CE enhancement (dB) |
| 0 | 0 | 0.912 | -48.4 | 0 | 0 | 0.912 | -48.4 | 0 |
| 1 | | | -40.8 | 7.6 | 3.4 | | -40.7 | 7.7 |
| 2 | | | -43.1 | 5.3 | 1.4 | | -41.0 | 7.4 |
| 3 | 3.71 | 0.912 | -49.1 | -0.7 | 0.6 | 0.912 | -41.8 | 6.6 |
| 4 | | | -54.9 | -6.5 | 0.4 | | -41.7 | 6.7 |
| 5 | | | -60.9 | -12.5 | 0.3 | | -42.4 | 6.0 |
| 10 | | | -45.9 | 2.5 | 0.081 | | -45.7 | 2.7 |
| 20 | | | -43.8 | 4.6 | 0.041 | | -43.9 | 4.5 |
| 30 | 0.05 | 0.912 | -42.5 | 5.9 | 0.029 | 0.912 | -41.7 | 6.7 |
| 40 | | | -40.7 | 7.7 | 0.021 | | -38.8 | 9.6 |
| 50 | | | -38.1 | 10.3 | 0.017 | | -34.7 | 13.7 |

[a] The experimental results are based on the measured values at $P_p = P_s = 22$ dBm in Ref. [18].

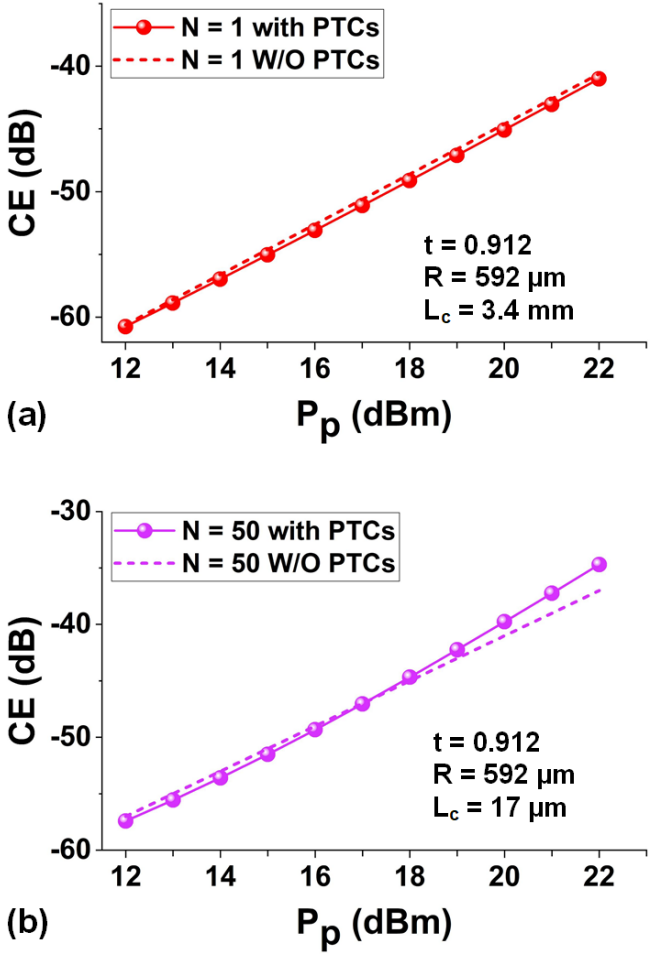

Fig. 7. CE comparison of GO-coated MRRs with and without (W/O) considering photo-thermal changes (PTCs) in GO films when (a) $N = 1$, $L_c =$ 3.4 mm and (b) $N = 50$, $L_c = 17$ µm. In (a) and (b), $t = 0.912$ and $R = 592$ µm.

## IV. OPTIMIZING FWM PERFORMANCE

In this section, we investigate the influence of the GO film parameters ($N$, $L_c$) and MRR coupling strength ($t$) on the FWM performance of the GO-coated MRRs, taking into account the power-dependent propagation loss and nonlinear parameter discussed in Section III. We assume the FWM test setup is the same as that in our previous experiment [18], and use MATLAB software to calculate the FWM CE based on classical FWM and MRR theory.

The FWM CE of the GO-coated MRRs ($CE_{MRR}$) is calculated by [23, 38]

$$CE_{MRR} = \frac{P_{idler,\,out}}{P_{signal,\,in}} = CE_{WG} \cdot FE_p^4 \cdot FE_s^2 \cdot FE_i^2 \quad (3)$$

where $P_{idler,\,out}$ and $P_{signal,\,in}$ are the output power of the idler and input power of the signal, respectively. $CE_{WG}$ is the CE of an equivalent waveguide with the same length as the circumference of the MRR. The calculation of $CE_{WG}$ is based on the theory in Refs. [13, 17]. For the MRRs with patterned GO films, $CE_{WG}$ is calculated by dividing the equivalent waveguides into coated and uncoated segments that have different propagation loss and nonlinear parameters. $FE_{p,s,i}$ in Eq. (3) are resonant field enhancement factors for the pump, signal, and idler, respectively, which can be expressed as [18, 22, 39]:

$$FE_{p,s,i} = \kappa \cdot t \,/\, [1 - t^2 \cdot A \cdot \exp(j \cdot \phi_{p,s,i})] \quad (4)$$

where $t$ and $\kappa$ are the field transmission and coupling coefficients defined in Table I, respectively. $\phi_{p,s,i}$ are the round-trip phase shift of the pump, signal, and idler, respectively, which can be given by:

$$\phi_{p,s,i} = k_{pu,su,iu}L_u + k_{pc,sc,ic}L_c \quad (5)$$

$k_{pc,sc,ic}$ and $k_{pu,su,iu}$ are the wavenumbers of the pump, signal, and idler for the GO coated and uncoated segments, respectively. Table II summarizes the process flow to calculate the FWM CE of the hybrid MRRs. On the basis of the fit propagation loss and nonlinear parameter in Figs. 3(c) and (d), five steps were repeated to calculate the CE of hybrid MRRs with different GO film parameters ($N$, $L_c$) and coupling strength ($t$).

We first analyze the FWM CE of the hybrid MRRs with a fixed coupling strength ($t$) but different GO film parameters ($N$, $L_c$). The calculated FWM CE versus coating length $L_c$ and input pump power $P_p$ is shown in Fig. 5, with (a-i) – (a-v) and (b-i) – (b-v) showing the results for $N = 1 - 5$ and $N = 10 - 50$, respectively. Similar to Fig. 4, a smaller range of $L_c$ is chosen for thicker films ($N \geq 10$) to clearly show the difference. To simplify the discussion, we used the same power for the pump and signal in our calculation, therefore $12 - 22$ dBm of $P_p$ in Fig. 5 corresponds to $15 - 25$ dBm of $P_{cw}$. In our calculation, we used constant $t = 0.912$ and $R = 592$ µm. The corresponding result for the uncoated MRR (when $L_c = 0$ mm) is also shown for comparison, which achieves the maximum CE of -48.4 dB at $P_p = P_s = 22$ dBm.

In Fig. 5, the CE of the hybrid MRRs first increases with GO film length $L_c$ and then decreases, achieving maximum values at intermediate film lengths. The optimized film length $L_c'$ that corresponds to the maximum CE decreases with $N$. This reflects the fact that the Kerr nonlinearity enhancement dominates for the devices with relatively small $L_c$ and $N$, and the influence of loss increase becomes more significant as $L_c$ and $N$ increase.

In Table III, we compare the calculated CE of the hybrid MRRs with optimized GO film lengths and the measured CE in our previous experiment where we fabricated devices with fixed film coating lengths of ~3.71 mm (i.e., the circumference of the MRR) for $N = 1 - 5$ and 50 µm for $N = 10 - 50$ [18]. For the devices with optimized GO film lengths, there is an improvement in the CE for all the considered GO layer numbers. Particularly, a maximum CE of -34.7 dB is achieved for $N = 50$ and $L_c = 17$ µm, which corresponds to a CE enhancement of 13.7 dB compared to the uncoated MRR and 3.4 dB further improvement relative to previous experimental result.

In addition to GO film parameters ($N$, $L_c$), the MRR coupling strength ($t$) also significantly affects the FWM performance of the hybrid MRRs. Based on the process flow in Table II, we further calculate the FWM CE of hybrid MRRs with different coupling strength ($t$). In our calculations, we chose 20 different values of $t$ ranging from 0.812 to 0.997. For each of them, the calculation processes for Fig. 5 with a fixed $t$ were repeated to obtain the optimized film length $L_c'$ and the corresponding maximum CE for different numbers of GO layers $N$.

Fig. 6(a) shows the calculated $L_c'$ versus $t$, (i) for $N = 1 – 5$ and (ii) for $N = 10 – 50$. The other device parameters are kept the same, i.e., $R = 592$ μm and $P_p = P_s = 22$ dBm. As can be seen, $L_c'$ decreases with $t$. This reflects the fact that the positive impact of the GO films in improving the FWM CE decreases as $t$ increases. For $N = 1$, $L_c'$ reaches the MRR's circumference (i.e., ~3.71 mm) at $t = 0.902$ and thus cannot be further increased for $t < 0.902$. Fig. 6(b) shows the maximum CE of the hybrid MRRs corresponding to the calculated $L_c'$ in Fig. 6(a). The results for the uncoated MRRs ($N = 0$) are also shown for comparison. The CE for $N = 10$ is lower than that for $N = 1 – 5$, mainly due to a more rapid increase of the propagation loss with $N$ than the nonlinear parameter $γ$. This can be attributed to increased imperfections, such as inhomogeneous defects, film unevenness, and imperfect contact between adjacent layers, for thicker GO films. The CE enhancement compared to the uncoated MRR is further extracted from Fig. 6(b) and plot in Fig. 6(c). A maximum CE enhancement of 18.6 dB is achieved at $t = 0.812$, $L_c = 42$ μm, and $N = 50$, which is 4.9 dB higher than the maximum CE enhancement when $t = 0.912$. This reflects the fact that reducing $t$ further yields a better CE enhancement. Here, we do not show the results for $t < 0.812$ mainly for two reasons. The first is due to the trade-off between achieving the maximum overall CE versus the maximum relative CE enhancement. As shown in Figs. 6(b) and (c), although reducing $t$ yields a better relative CE enhancement, it also results in a lower CE. For example, for $t < 0.812$, the maximum CE is $< -40$ dB. The second reason is because the $FE_{p, s, i}$ factors in Eq. (4) decrease with $t$. For $t < 0.812$, the $FE_{p, s, i}$ are close to 1. This results in a $CE_{MRR}$ close to $CE_{WG}$ in Eq. (3), indicating that there is little CE improvement induced by resonance enhancement of the MRR. The difference in CE between the hybrid and uncoated MRRs becomes smaller as $t$ increases, which is consistent with the trend for $L_c'$ in Fig. 6(a). When $t$ is close to 1, the CE enhancement approaches zero, indicating that incorporating GO films would not bring any benefits in improving the FWM performance in this case. In Fig. 6(d), we plot the insertion loss (at the drop port) of the hybrid MRRs with optimized film lengths $L_c'$ in Fig. 6(a). It can be seen that the insertion loss increases with $t$ and becomes $> 8$ dB when $t$ is close to 1, which is mainly induced by the four-port MRRs with two identical directional couplers. This indicates that despite the MRR with a weak coupling strength (i.e., high $t$) has a high CE, it suffers from a high insertion loss that limits their use in practical applications.

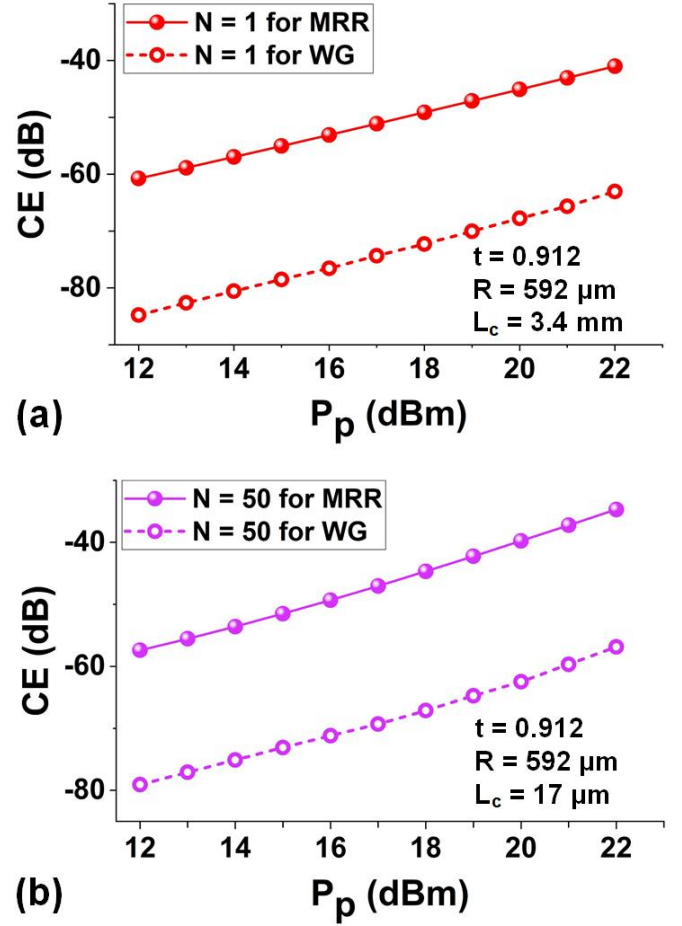

Fig. 8. CE comparison of GO-coated MRRs and comparable GO-coated waveguides when (a) $N = 1$, $L_c = 3.4$ mm and (b) $N = 50$, $L_c = 17$ μm. In (a) and (b), $t = 0.912$ and $R = 592$ μm. WG: waveguide.

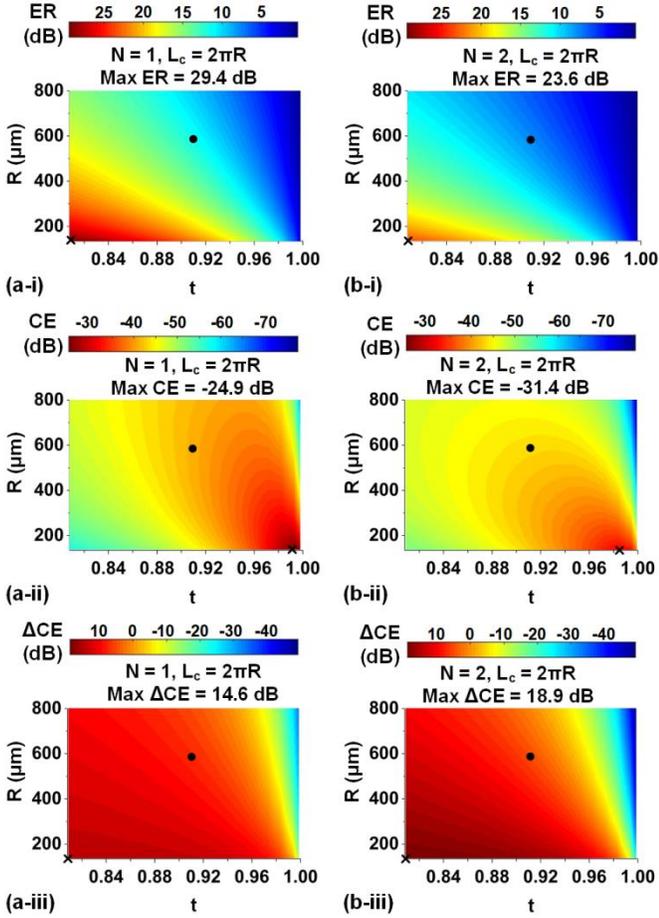

Fig. 9. Performance comparison of uniformly GO-coated MRRs when (a) $N$ = 1 and (b) $N$ = 2. In (a) – (b), $P_p = P_s$ = 22 dBm, (i) − (iii) show the ER, CE, and CE enhancement versus $R$ and $t$, respectively. The black circles mark the experimental results in Ref. [18] and the black crosses mark the maximum values in each figure. ΔCE: CE enhancement compared to uncoated MRRs with the same $t$ and $R$.

## V. DISCUSSION

In this section, we discuss the influence of photo-thermal changes in the GO films as well as the effect of varying some of the other MRR paramteres such as ring radius and waveguide dispersion on the FWM performance. This, together with the analysis in Section IV, provides a systematic approach for designing GO-coated MRRs in order to optmize the FWM performance.

As discussed in Section III, photo-thermal changes in the GO films lead to power-dependent propagation loss and nonlinear parameter $\gamma$ for the hybrid MRRs. Both of these parameters affect the FWM CE and there is a trade-off between them. In Fig. 7, we compare the FWM CE of the hybrid MRRs with and without considering any photo-thermal changes. For the hybrid MRRs without any photo-thermal changes, we used a constant propagation loss and nonlinear parameter, equivalent to their values at low powers. In Figs. 7(a) and (b), we show the results for the hybrid MRRs with 1 and 50 layers of GO. For each of them, optimized film lengths were chosen. The other device parameters are kept the same, i.e., $t$ = 0.912 and $R$ = 592 μm. In Fig. 7(a), after including photo-thermal changes, the CE decreases, with a more notable difference occurring at higher powers. This reflects the fact that the influence of an increase in loss is more significant than the increase of $\gamma$ for the device with a thin GO film. In Fig. 7(b), the CE obtained when including photo-thermal effects is lower at low pump powers, while as the pump power increases, it gradually overtakes the CE obtained without including photo-thermal effects. This reflects a more complex influence of the photo-thermal changes on the FWM performance for the hybrid MRRs with thick GO films, which can be attributed to an increase of defects and imperfect contact as well as more obvious thermal dissipation issue in the thick GO films.

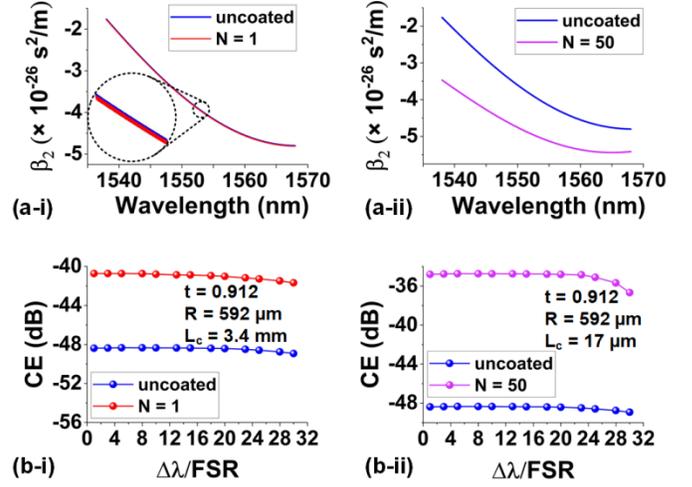

Fig. 10. (a) Group-velocity dispersion $\beta_2$ for hybrid MRRs with (i) $N$ = 1 and (ii) $N$ = 50 layers of GO. (b) CE versus Δλ/FSR for hybrid MRRs when (i) $N$ = 1, $L_c$ = 3.4 mm and (ii) $N$ = 1, $L_c$ = 17 μm. In (a) and (b), $t$ = 0.912, $R$ = 592 μm, and $P_p = P_s$ = 22 dBm. The corresponding results for the uncoated MRR are also shown for comparison.

Due to the resonant enhancement effect in the MRRs, the FWM CE can be significantly improved in GO-coated MRRs as compared with GO-coated waveguides. In Fig. 8, we compare the FWM CE of GO-coated MRRs and comparable GO-coated waveguides, (i) for the devices with 1 layer of GO and (ii) for the devices with 50 layers of GO. Similar to the case of Fig. 7, optimized film lengths were chosen for the hybrid MRRs and the other device parameters are kept the same as those in Fig. 7. The hybrid waveguides have the same length as the circumference of the MRRs, and both the MRR and the waveguides have the same GO film length. For the hybrid waveguides, we neglect the slight variation induced by photo-thermal changes in the GO films. As can be seen, the CEs of the hybrid MRRs are much higher than those of the hybrid waveguides for both $N$ = 1 and $N$ = 50, clearly reflecting the huge CE improvement enabled by the resonant structure.

For practical device fabrication, hybrid MRRs with uniformly coated GO films are easier to be fabricated since they do not need lithography or lift-off processes for film patterning. In Fig. 9, we further investigate the FWM performance of these hybrid MRRs. Fig. 9(a-i) shows the MRR's ER versus its radius $R$ and coupling strength $t$ when $N = 1$ and $P_p = P_s = 22$ dBm. The ER decreases with both $R$ and $t$ – the former results from the increase of the intracavity loss with $R$, while the latter is consistent with the trend in Fig. 2(b). Fig. 9(a-ii) shows the CE versus $R$ and $t$. The CE enhancement relative to the uncoated MRR is further extracted from Fig. 9(a-ii) and shown in Fig. 9(a-iii). In our calculation, we neglect the slight difference in the MRR coupling strength $t$ between the GO coated and uncoated MRRs, since including it would result in a difference of only < 0.3%. In Fig. 9(a-ii), the CE (-40.8 dB) at $R = 592$ µm and $t = 0.912$ is marked, which corresponds to a CE enhancement of 7.6 dB in Fig. 9(a-iii), showing good agreement with the experimental result in Ref. [18]. The maximum CE (-24.9 dB) at $R = 135$ µm and $t = 0.992$ is also marked, which is 15.9 dB higher than the CE at $R = 592$ µm and $t = 0.912$ and corresponds to a CE enhancement of -1.8 dB. In Fig. 9(a-iii), a maximum CE enhancement of 14.6 dB is achieved at $R = 135$ µm and $t = 0.812$, which is different to the point corresponding to the maximum CE. This reflects the trade-off between achieving the maximum CE versus the maximum relative CE enhancement for the device design, which is consistent with the results in Figs. 6(b) and (c). Fig. 9(b) shows the corresponding results for $N = 2$. The maximum CE enhancement is improved further by ~4.3 dB as compared with that for $N = 1$, while both the maximum ER and CE decrease due to the increase in loss with film thickness. This, on one hand, indicates that a high CE enhancement can be achieved for the hybrid MRRs with small radii even without the use of film patterning, while on the other hand, it reflects the fact that the CE significantly decreases with GO film thickness for the uniformly coated MRRs.

Finally, we investigate the influence of waveguide dispersion on the FWM performance of hybrid MRRs. Fig. 10(a) shows the group-velocity dispersion $\beta_2$ for the hybrid MRRs with (i) $N = 1$ and (ii) $N = 50$ layers of GO, together with the $\beta_2$ of the uncoated MRR. The material dispersion of GO and doped silica was taken from Refs. [12, 27]. The $\beta_2$ of the hybrid MRRs is slightly lower as compared with the uncoated MRR, with the difference becoming more significant for the device with thicker films. The reduced $\beta_2$ induced by the GO films yields an enhanced anomalous dispersion and consequently better phase matching for FWM [40]. Fig. 10(b) shows the CE versus $\Delta\lambda$ (defined as wavelength spacing between pump and signal) for the hybrid MRRs, (i) for $N = 1$, $L_c = 3.4$ mm and (ii) for $N = 50$, $L_c = 17$ µm. The corresponding result for the uncoated MRR is also shown for comparison. The other parameters are kept the same as $t = 0.912$ and $R = 592$ µm. The CE slightly decreases with $\Delta\lambda$, with a difference < 2 dB for $\Delta\lambda$ / FSR = 30 when $N = 50$, $L_c = 17$ µm. This reflects the fact that both the doped silica and the GO film have a low material dispersion, which allows highly effective phase matching for broadband FWM.

VI. CONCLUSION

In summary, the FWM performance of MRRs integrated with 2D layered GO films is theoretically studied and optimized based on material and device parameters from previous experiments. A detailed analysis for the influence of GO film parameters and MRR coupling strength on the FWM CE of the GO-coated MRRs is performed. By redesigning the device parameters to properly balance the trade-off between the Kerr nonlinearity and loss, up to ~18.6 dB enhancement in the FWM CE is achieved, corresponding to ~8.3 dB further improvement over what was achieved experimentally. The influence of photo-thermal changes in the GO films as well as the variation of some other MRR parameters such as ring radius and waveguide dispersion is also investigated. These results confirm the effectiveness of introducing GO films to improve the MRR's FWM performance and serve as a roadmap for optimizing the FWM performance of GO-coated MRRs.